\documentclass[preprint,showpacs,preprintnumbers,amsmath,amssymb]{revtex4}

\usepackage{graphicx}
\usepackage{dcolumn}
\usepackage{bm}

\begin{document}
\title{Tunable 0.7 conductance plateau in quantum dots}
\author{Yunchul Chung$^{1}$}
\author{Sanghyun Jo$^{2}$}
\author{Dong-In Chang$^{2}$}
\author{Hu-Jong Lee$^{2}$}
\email{hjlee@postech.ac.kr}
\author{M. Zaffalon$^{3}$}
\author{V. Umansky$^{3}$}
\author{M. Heiblum$^{3}$}

\affiliation{$^1$Department of Physics, Pusan National
University, Busan 609-735, Republic of Korea \\
$^2$Department of Physics, Pohang University of Science and
Technology, Pohang 790-784, Republic of Korea\\
National Center for Nanomaterials Technology, Pohang 790-784,
Republic of Korea\\
$^3$Department of Physics, Weizmann Institute of Science, Rehovot
76100, Israel}
\date{\today}

\begin{abstract}
A new consistent approach to forming the 0.7 structure by using a
quantum dot rather than a quantum point contact is demonstrated.
With this new scheme, it was possible to tune on and off the 0.7
structure. The new 0.7 structure continuously evolved into a
normal integer conductance plateau by varying the tuning
condition. Unlike the conventional 0.7 plateau, the new 0.7
structure was observed even at low electron temperatures down to
100 mK, with unprecedented flatness. From our results, it is
concluded that electron interference as well as the electron
interaction effect should be taken into consideration to explain
the 0.7 structure.
\end{abstract}

\pacs{71.70.Ej, 73.23.Ad, 73.21.La}

\maketitle

The quantization of conductance in units of $G_0$ $(=2e^2/h)$
through a quantum point contact (QPC) was first observed
experimentally by two groups \cite{Wees,Pepper}. The phenomenon is
a result of adiabatic transmission of electrons through
spin-degenerate non-interacting one-dimensional (1D) channels.
Shortly after the discovery, an additional plateau was observed at
around $0.7G_0$, which has been known as the {\it 0.7 structure}
\cite{Pepper first}. Many other experiments were followed to
confirm the existence of this unusual plateau and to investigate
its physical origin \cite{Pepper1, Pepper2,Pepper3,Marcus
Cronenwett,Pfeiffer, Glattli,Rafi}. Since this anomalous
conductance plateau cannot be explained in a single-particle
picture, theoretical explanations based on electron-electron
interaction, spin effect, and others have been suggested
\cite{Reilly, Havu, Berggren, Meir, Oreg}. The spontaneous spin
polarization model \cite{Reilly,Havu} and the Kondo-related model
\cite{Meir} are most popular among them, which were supported
experimentally by Thomas {\it et al.} \cite{Pepper first} and
Cronenwett {\it et al.} \cite{Marcus Cronenwett}, respectively.

Nonetheless, no consensus has been reached about its origin yet
and still contradicting experimental and theoretical results are
reported. Furthermore, the 0.7 structure is not always revealed in
a QPC although the phenomenon is regarded as intrinsic properties
of 1D conduction channels. In most studies, however, the absence
of the 0.7 structure was simply ignored and the condition for the
occurrence of this structure has not been systematically
investigated. We report a new approach to the formation of the
anomalous 0.7 structure by using a quantum-dot structure, which
allowed us to tune to the optimal observing condition at our
disposal. This tunable 0.7 structure showed characteristics that
were similar to the conventional ones in a QPC, but with
additional new features. We believe our results will provide an
insight into the nature of this ever-controversial phenomenon.

The quantum dots were fabricated on two-dimensional electron gas
(2DEG) wafers based on a GaAs/AlGaAS heterostructure. The electron
densities were around $2.5 \times 10^{11}$ cm$^{-2}$ with the
mobilities higher than $1.5 \times 10^6$ cm$^2$V$^{-1}$s$^{-1}$ at
4.2 K. The 2DEG layer resided 65 nm below the surface of the
wafers. Two different types of quantum dots (Types A and B) were
fabricated (Fig. 1). The electron temperature was estimated to be
around 100 mK (140 mK) for the Type-A (Type-B) sample. For
convenience, all the gates were named individually as in Figs.
1(a) and 1(b). Conductance quantization behavior for individual
QPC's was examined by applying negative voltages on the nose gate
and one of the other three gates (QPC, side, and plunger gates)
separately, while keeping the rest of the two gates at zero
voltage. In each of these usual QPC geometries, only the normal
integer conductance quantization was observed without any
anomalies including the 0.7 structure. We believe the absence of
anomalies was caused by the extreme narrowness (less than 50 nm in
its width) of our gates. The zero-bias anomaly suggesting the
Kondo character \cite{Marcus Cronenwett} was not observed, either,
even at 100 mK, the base electron temperature used.

As in Fig. 1(c), the conductance anomaly started to appear only
when certain negative voltages were applied to all the gates
concurrently. To induce the conductance anomaly we first applied
small fixed negative voltages to the QPC and the plunger gates,
which were just strong enough to pinch the 2DEG under the gate.
Only the nose gate was then varied to observe conductance
quantization. For a fixed nose-gate voltage $V_{ng}$, the
side-gate voltage $V_{sg}$ was scanned from -0.15 to -0.6 V to
tune on and off the conductance anomaly. Fig. 1(c) shows the
conductance variation for the Type-B sample at the electron
temperature of 140 mK. The 0.7 structures are seen for $V_{sg}$
around -0.46 V and -0.34 V. Thus, controlling $V_{sg}$ in the
above circumstances provides a very convenient means to tune to
the formation of the 0.7 structure. Higher-order anomalies like
1.7 and 2.7 structures, observed in ordinary QPC's, are also
revealed. Similar features were obtained in the Type-A sample.

Figure 2(a) shows in detail how the 0.7 plateau for the Type-B
sample evolves to and from the normal integer plateaus as $V_{sg}$
varies between -0.24 V and -0.41 V. The 0.7 plateau becomes most
pronounced for $V_{sg}$=-0.34 V (the thick curve). For
$V_{sg}$=-0.24 V and above, however, no appreciable conductance
anomalies were observed (see Fig. 1). The width of $G_0$ plateau
reduces as a more negative voltage is applied to the side gate.
For $V_{sg}$=-0.34 V, it reduces to almost one half of that for
-0.24 V, with the concurrent development of a clear plateau just
below $0.8G_0$. Further increasing the negative voltage to the
side gate, the plateau reduces to a broad local maximum and
transforms back to the $G_0$ plateau for $V_{sg}$=-0.41 V. As
shown in Fig. 1(c) these generic processes repeat for varying
$V_{sg}$. The width of the $G_0$ plateau changes noticeably with
$V_{sg}$ varying from -0.24 V to -0.41 V, while the pinch-off
voltage remains almost unaltered. This implies that the recurrent
change of the width of the normal integer plateaus for varying
$V_{sg}$ was not caused by the possible electrostatic coupling
between the side and the QPC gates.

Figure 2(b) shows the modulation of the width of the $G_0$ plateau
for the Type-B sample as a function of $V_{sg}$. The strong
modulations are shown for $V_{sg}$ lower than -0.3 V, along with
clear 0.7 plateaus (two front arrows). At $V_{sg}$=-0.22 V a faint
0.7-like structure (the rear arrow) is also evident. Both samples
show well-formed plateaus for the conductance just below $0.8G_0$.
For the Type-A sample, with the average spacing between the side
and the QPC gates closer than in the Type-B sample, the plateau is
substantially flatter. Figure 2(c) shows the flatness of the 0.7
structure of the Type-A sample. The average conductance for
$V_{ng}$ between -1.120 V and -1.134 V is $0.791G_0$, with the
conductance deviation less than 0.1\% of the averaged value. No
0.7 structures reported to date have the flatness as high as this.
It is highly unlikely that such high flatness is induced by the
suppression of the Kondo resonance peak \cite{Meir}. The flatness
can hardly be explained without introducing an energy gap in the
model. Along this line, the spin-gap model \cite{Reilly, Havu} may
be the best candidate to interpret the data. It, however, does not
explain the repeated occurrence of the 0.7 structure for varying
$V_{sg}$. According to the model, the spin gap widens
monotonically with increasing the electron carrier density in a
QPC. The electron carrier density inside our QPC gate supposedly
decreased monotonically for more negative $V_{sg}$, the trend of
which was in contradiction to the clear occurrence of the 0.7
structure.

In this study the electron carriers were intended to be weakly
localized inside the quantum dot formed by the carrier-confining
gates, mimicking the Kondo \cite{Kondo} configuration that was
claimed to cause the 0.7 structure \cite{Meir}. The side-gate
voltages in the presence of the 0.7 structure [denoted by two
front arrows in Fig. 2(b)], $V_{sg}$=-0.34 V and -0.46 V,
corresponded to opening of 5.9 and 3.1 conducting channels between
the nose and the side gates, respectively. Thus, the carriers were
not effectively confined inside the quantum dot even in the
presence of the 0.7 structure. A weak 0.7 structure [denoted by
the rear arrow in Fig. 2(b)] was also observed even for much less
confining side-gate voltage, $V_{sg}$=-0.22 V, which corresponded
to opening of more than 100 conducting channels. This analysis,
thus, excludes the possibility of the Kondo origin of the 0.7
structure at least in our sample configuration.

The temperature dependence of the 0.7 structure of the Type-A
sample is shown in Fig. 3. Before varying the temperature, the
side gate was set at the optimum voltage ($V_{sg}=-0.19$ V) for
the formation of the 0.7 structure at the base electron
temperature. As shown in Fig. 3(a) the tunable 0.7 plateau gets
weaker as the temperature increases and eventually vanishes for
$T$=800 mK. The temperature dependence of the conductance in Fig.
3(b), similar to the one in Fig. 3(a) but taken for a slightly
more negative side-gate voltage ($V_{sg}=-0.20$ V), reveals a
local conductance maximum rather than a plateau at the base
electron temperature. As the temperature increases the local
maximum gets smeared and gradually transforms into the 0.7
structure, forming a perfect plateau at the temperature around 520
mK. This temperature-dependent evolution of the plateau is similar
to the feature of the regular 0.7 structure observed in the
conventional QPC's \cite{Pepper first, Marcus Cronenwett}, where
it has been known that the 0.7 structure develops only at
relatively high temperatures around 1 K.

The source-drain bias dependence of the conductance at different
$V_{ng}$ was examined with $V_{sg}$ set to show an almost normal
integer conductance plateau. As seen in Fig. 4(a), the measured
conductance curves around zero bias are somewhat asymmetric, due
to the imbalance of the potential build-up around each gate. The
0.7 structures are seen for the bias around 0.87 mV, 0.46 mV,
-0.36 mV, and -0.87 mV. One also notices that the 0.7 structure is
a little clearer in negative biases, which correspond to the
situation where electrons were injected to the side gate first.
Since the side gate was more open than the QPC gate, the carrier
density in this case was higher in the quantum-dot region between
the side and the QPC gates than the case of electrons injected to
the QPC gate first. If the new 0.7 structure observed in our
samples was caused by some correlation of electrons in the
carrier-confining quantum-dot region, a stronger effect was more
likely to be observed when the electron concentration was higher
as in this study.

Measurements similar to the ones leading to Fig. 1(c) were
repeated in a magnetic field of 0.15 T applied perpendicular to
the plane of the 2DEG. As shown in Fig. 4(b), in this relatively
low magnetic field, the 0.7 structures were almost entirely
suppressed, leaving only normal integer conductance plateaus. A
very weak field like 0.05 T was high enough to considerably weaken
the 0.7 structures. No such behavior has been reported in the
conventional 0.7 conductance anomaly \cite{mag effect}.

With two carrier-confining gates (the side and the QPC) arranged
in series in the direction of electron transport, the interference
may have taken place for the electrons moving back and forth
between the gates. It thus can be inferred that such an
interference may have caused the 0.7 plateau in our measurements.
In Fig. 4(c), the conductance is plotted as a function of
$V_{sg}$, with the nose gate fixed around the middle of the 0.7
plateau. A clear interference pattern is revealed as $V_{sg}$
varies. Marked by arrows in Fig. 4(c) are the positions of
$V_{sg}$ where the 0.7 structures were observed. Although the
interference is evident, not much correlation exists between the
0.7 structure and the interference pattern. The 0.7 structure
formed at $V_{sg}$=-0.22 V locates at the local conductance
maximum, while those formed for $V_{sg}$ at -0.46 V and -0.34
locate somewhat away from the local conductance maxima. This
implies that the 0.7 structure was not caused by the interference
only. It is possible that multiple transmission modes participated
in the interference. If that was the case no direct correlation
could exist between the 0.7 structure and the interference pattern
as in Fig. 4(c).

The extreme flatness of the 0.7 structure over a wide range of
$V_{ng}$ as in Fig. 2(c) is not explained by the simple electron
interference effect, either. Varying the $V_{ng}$, which is bound
to alter the phase of the interfering electrons, should have
resulted in the continuous change in the conductance as a function
of $V_{ng}$. In addition, the observation of the 0.7 plateau at
high temperatures shown in Fig. 3(b) is hardly explained by the
interference. At such high temperatures the interference is
supposed to be much reduced due to the decoherence of electron
phases. On the other hand, the repetition of 0.7 structure as a
function of side-gate and bias voltages resembles the feature
often seen in the electron interference effect.

Recently, it has been reported that a 0.5 plateau can be realized
by changing the potential landscape of 1D wire by using a scanning
probe tip \cite{Pepper4}. The 0.5 structure was observed when the
potential away from the main confinement potential landscape was
perturbed. In this case, electrons scattering back and forth
between the main and perturbed potential possibly resulted in the
interference. Even in conventional QPC's the electron interference
is likely to happen since the shape of the quasi-1D wire formed by
QPC gates is usually distorted by the formation of the unintended
impurity potential. An exception takes place for an extremely
short QPC formed by narrow gates. In this study, no conductance
anomaly was observed for QPC's, the widths of which were shorter
than 50 nm. We believe that the 0.7 structure has not been
observed consistently in ordinary QPC's due to the varying
distortion of the potential landscape from sample to sample. The
difference in features of the 0.7 structure between our samples
and conventional QPC's can be a result of the difference in the
detailed shape of the potential landscape. Although we
consider the electron interference is 
the main causes of the 0.7 anomaly, all of its features are not
explicable in terms of the electron interference only. Thus, both
the interference and the interaction effect are believed to play
essential roles in causing the 0.7 anomaly.

In summary, we have demonstrated a new approach to forming the 0.7
plateau in tunable quantum-dot structures. In clear contrast to
the conventional one in an ordinary QPC structure, the 0.7 plateau
in our study was observed even at very low electron temperatures
with unprecedented flatness. This new 0.7 plateau evolved into a
normal integer plateau when the negative side-gate voltage was
varied. Since the 0.7 plateau occurs even when carriers are not
fully confined in the quantum-dot region the Kondo mechanism is
ruled out for the cause of the anomaly at least in our tunable
quantum-dot structures. Interference as well as enhanced
interaction among electrons in and around a QPC is believed to be
the main cause of the anomaly. This new approach allows one to
tune to the formation condition of the 0.7 plateau at one's
disposal. The tuning, however, turns out to be somewhat subtle,
which may explain the reason the conventional 0.7 structure is not
always observed in ordinary QPC's.

This work was supported by Electron Spin Science Center, in Pohang
University of Science and Technology, administered by KOSEF, and
by the Korea Research Foundation Grant, KRF-2005-070-C00055 and
KRF-2004-003C00076.

\newpage
\textbf{FIGURE CAPTIONS}
\\

Figure 1. SEM pictures of two different types of quantum dots; (a)
Type A and (b) Type B. (c) The conductance measured for the Type-B
sample, with the QPC-gate and the plunger-gate voltages fixed at
-0.7 V and -0.23 V, respectively, as a function of nose-gate
voltages $V_{ng}$ while varying the side-gate voltages $V_{sg}$
from -0.15 V to -0.6 V at intervals of 0.01 V. The 0.7 structure
is present at $V_{sg}$=-0.46 V and -0.34 V (lower solid curves).
Weak 0.7 anomaly is also seen at $V_{sg}$=-0.22 V (upper solid
curve). Data for $V_{sg}$ from -0.24 V to -0.41 V are shown in
more detail in Figure 2.
\\
\\

Figure 2. (color) (a) The evolution of the 0.7 plateau of the
Type-B sample to and from the normal integer plateaus for varying
$V_{sg}$ from -0.24 V to -0.41 V at intervals of 0.01 V. The thick
curve is the conductance taken for -0.34V. (b) The conductance as
a function of $V_{sg}$ and $V_{ng}$ put in a 3D-pseudo-color plot.
The positions of the 0.7 structure are marked by arrows for
clarity. (c) The conductance as a function of the $V_{ng}$ for the
Type-A sample at the electron temperature of 100 mK, with the side
and the QPC gates were set at -0.19 V and -0.85 V, respectively.
\\
\\

Figure 3. (color online)  The temperature dependence of 0.7
structure for the Type-A sample. The nose and the QPC gates were
set to show (a) the 0.7 structure and (b) a resonance-like local
conductance maximum  rather than a plateau at 92 mK. The fridge
temperatures were 92, 152, 250, 365, 420, 520, 630, 730, and 830
mK from bottom to top [without the data for 630 mK in (b)].
\\
\\

Figure 4. (a) The differential conductance of the Type-B sample,
measured as a function of the source-drain bias voltage for
varying $V_{ng}$ from -632 mV to -1061 mV at intervals of 3 mV.
The QPC and the plunger gates were set at -0.7 V and -0.23 V,
respectively, while the side gate was at -0.39 V. (b) The tunable
0.7 structure taken in the same measurement condition as the data
in Fig. 1(c) but in a perpendicular magnetic field of 0.15 T. (c)
The conductance as a function of $V_{sg}$ with $V_{ng}$ set at
-0.964 V. The 0.7 structures were observed at $V_{sg}$ marked by
the arrows.

\newpage

\begin{figure}[p]
\begin{center}
\vspace{3cm}\noindent\hspace{.5cm}
\scalebox{0.85}{\includegraphics{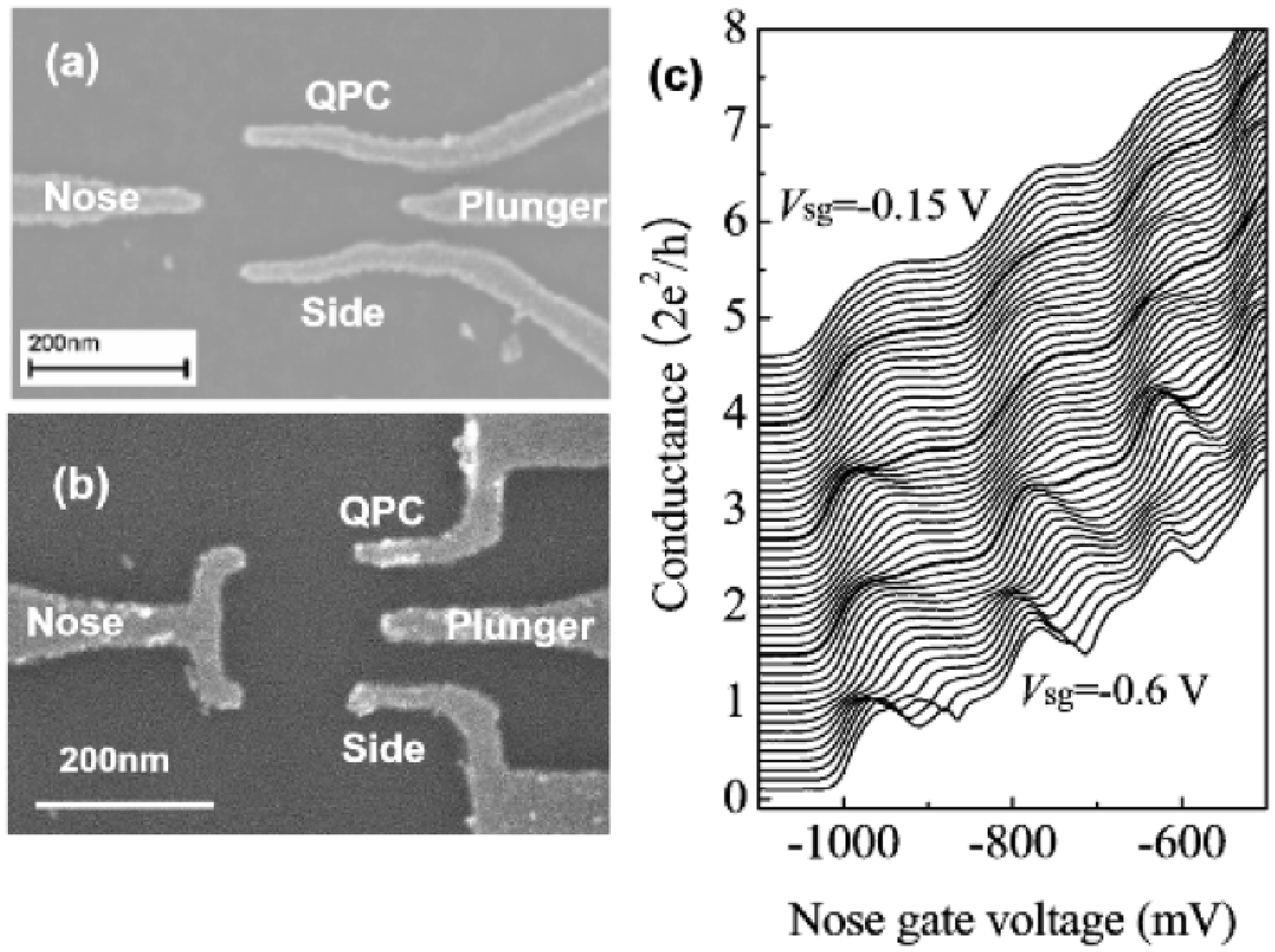}} \vspace{0.5cm} \caption{}
\end{center}
\end{figure}

\newpage

\begin{figure}[p]
\begin{center}
\vspace{-26cm}\noindent\hspace{.2cm}
\scalebox{0.9}{\includegraphics{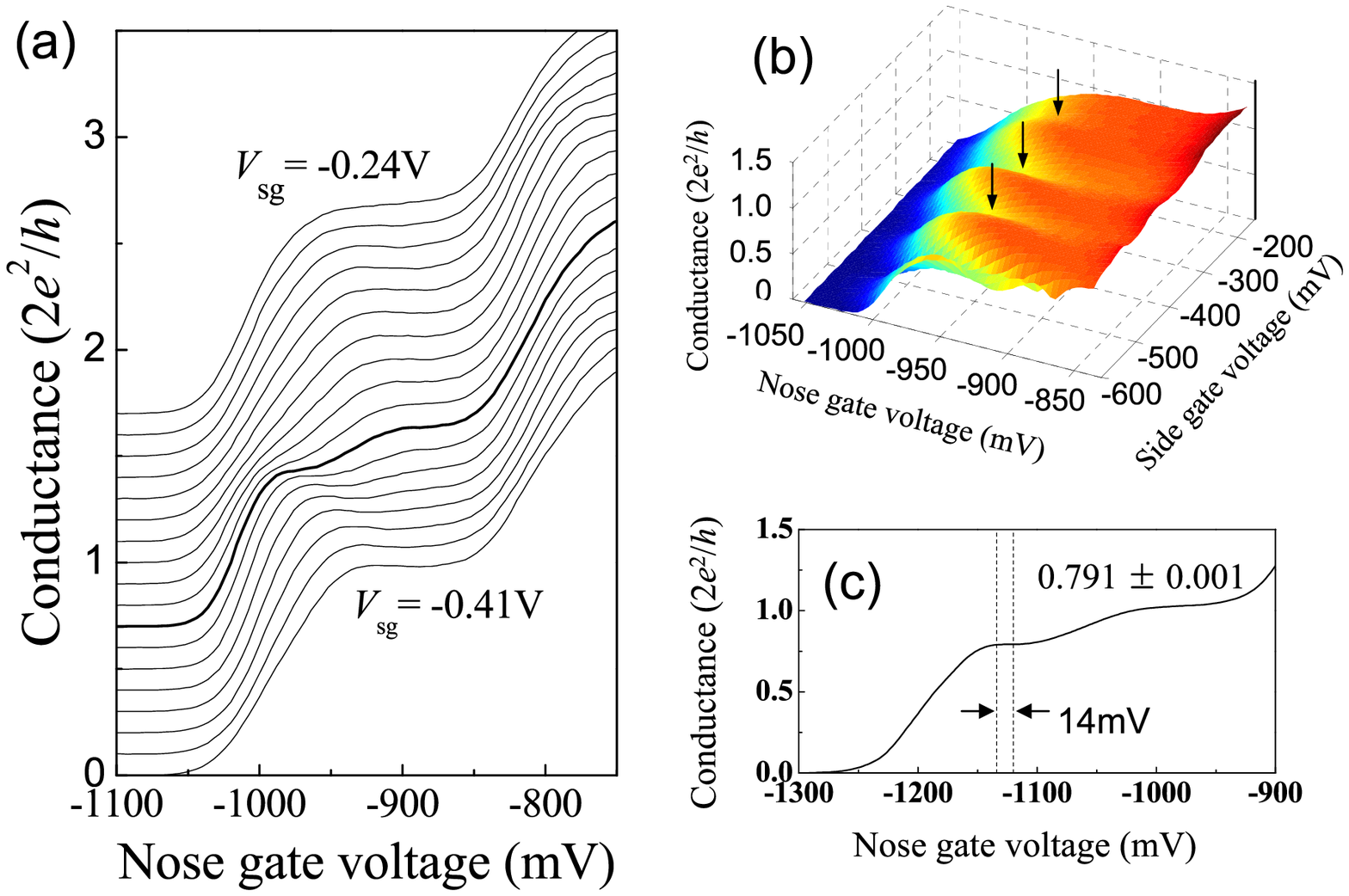}}
\vspace{-4.5cm}
\caption{}
\end{center}
\end{figure}

\newpage

\begin{figure}[p]
\begin{center}
\vspace{-26cm}\noindent\hspace{0cm}
\scalebox{.95}{\includegraphics{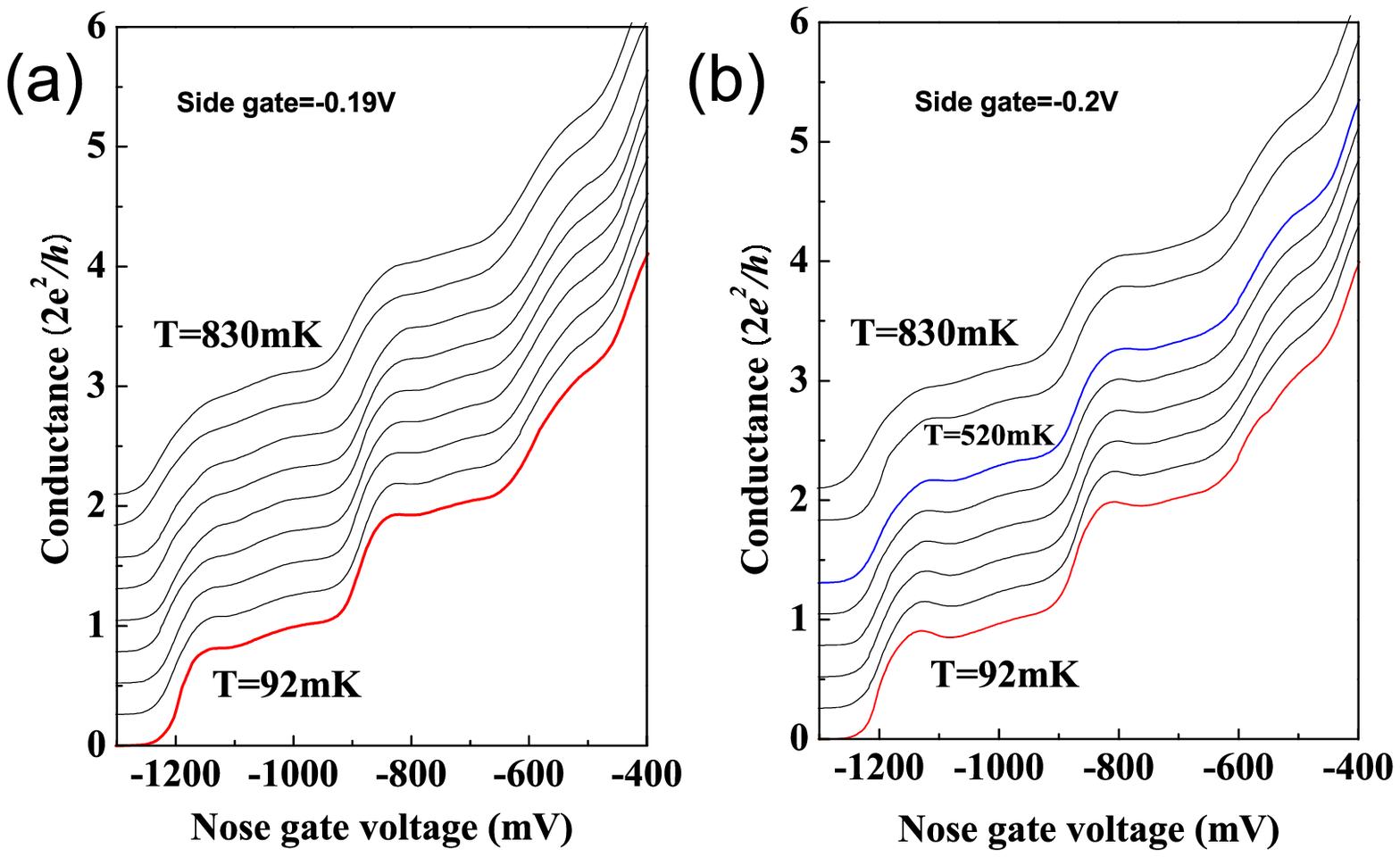}}
\vspace{-4.5cm}
\caption{}
\end{center}
\end{figure}

\newpage

\begin{figure}[p]
\begin{center}
\vspace{-25cm}\noindent\hspace{0cm}
\scalebox{.9}{\includegraphics{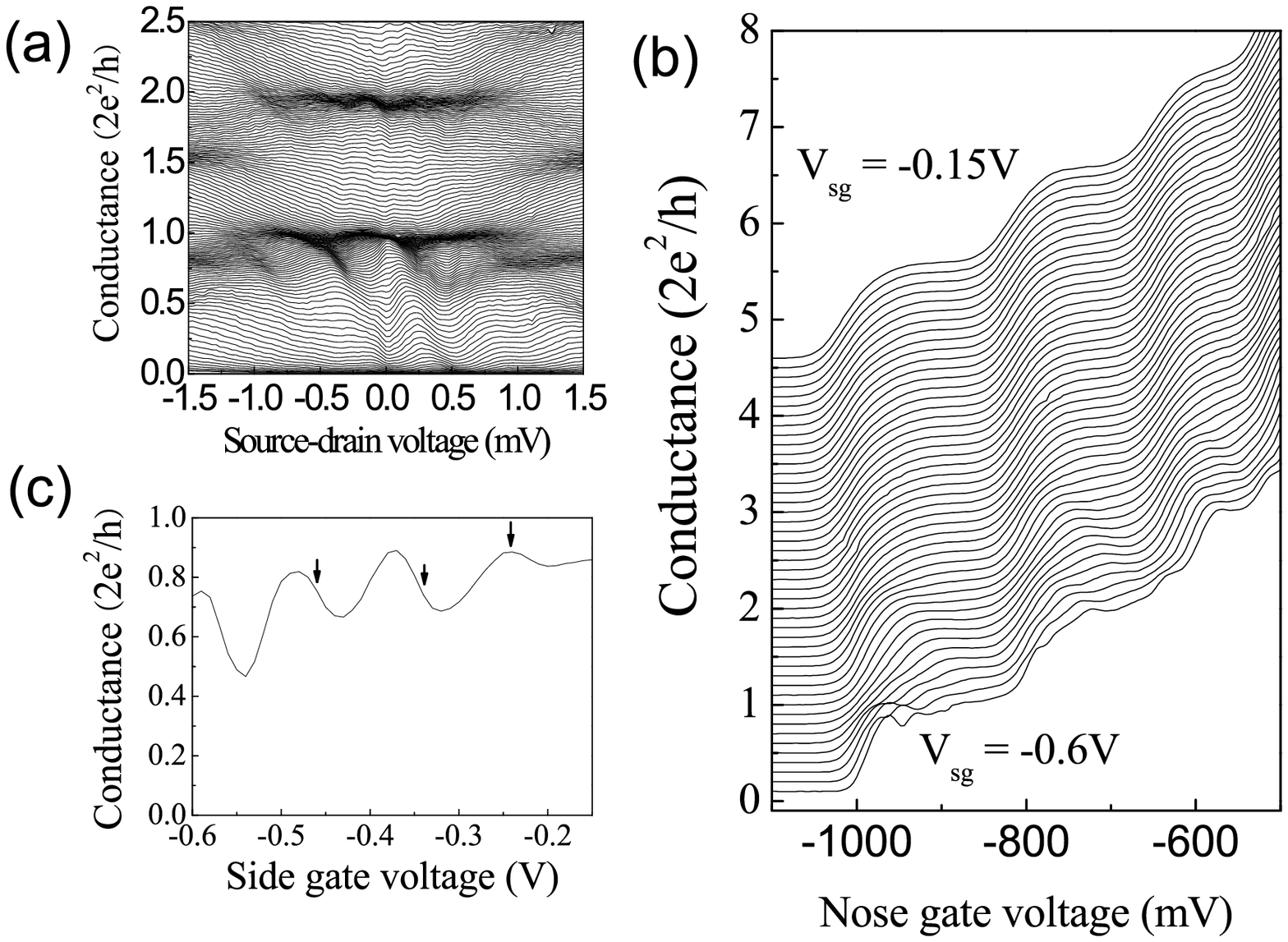}}
\vspace{-4.5cm}
\caption{}
\end{center}
\end{figure}


\begin{thebibliography}{00}
\bibitem{mag effect} The 0.7 structure was also examined for fields applied in parallel
with the 2DEG. The transverse magnetoresistance measurements could
not be made, however, because the perpendicular field component
caused by any possible field misalignment from the planar
direction quickly suppressed the 0.7 structure.
\end{thebibliography}
\end{document}